\documentclass[aps,pre,twocolumn,showpacs,showkeys,groupedaddress,floatfix]{revtex4-1}
\usepackage{graphicx}
\usepackage{amsmath}
\usepackage{amssymb}

\begin{document}

\title {Jamming and percolation in generalized models of random sequential adsorption of linear $k$-mers on a square lattice}
\author{Nikolai I. Lebovka}
\email[Correspondence author: ]{lebovka@gmail.com}
\affiliation{Institute of Biocolloidal Chemistry named after F.D. Ovcharenko, NAS of Ukraine, Kiev, Ukraine}
\affiliation{Taras Shevchenko Kiev National University, Department of Physics, Kiev, Ukraine}
\author{Yuri Yu. Tarasevich}
\email[Correspondence author: ]{tarasevich@asu.edu.ru}
\affiliation{Astrakhan State University, Astrakhan, Russia}
\author{Dmitri O. Dubinin}
\affiliation{Astrakhan State University, Astrakhan, Russia}
\author{Valeri  V. Laptev}
\affiliation{Astrakhan State University, Astrakhan, Russia}
\affiliation{Astrakhan State Technical University, Astrakhan, Russia}
\author{Nikolai V. Vygornitskii}
\affiliation{Institute of Biocolloidal Chemistry named after F.D. Ovcharenko, NAS of Ukraine, Kiev, Ukraine}

\date{\today}

\begin{abstract}
The jamming and percolation for two generalized models of random sequential adsorption (RSA) of
linear $k$-mers (particles occupying $k$ adjacent sites) on a square lattice are studied by means of Monte Carlo simulation. The classical random sequential adsorption (RSA) model assumes the absence of overlapping of the new incoming particle with the previously deposited ones.
The first model LK$_d$ is a generalized variant of the RSA model for both $k$-mers and a lattice with defects. Some of the occupying $k$ adjacent sites are considered as insulating and some of the lattice sites are occupied by defects (impurities). For this model even a small concentration of defects can inhibit percolation for relatively long $k$-mers.
The second model is the cooperative sequential adsorption (CSA) one, where, for each new $k$-mer, only a restricted number of lateral contacts $z$ with previously deposited $k$-mers is allowed. Deposition occurs in the case when $z\leq (1-d)z_m$  where $z_m=2(k+1)$ is the maximum numbers of the contacts of $k$-mer, and $d$ is the fraction of forbidden NN contacts. Percolation is observed only at some interval $k_{min}\leq k\leq k_{max}$ where the values  $k_{min}$ and $k_{max}$ depend upon the fraction of forbidden contacts $d$. The value $k_{max}$  decreases as $d$ increases.  A logarithmic dependence of the type $\log(k_{max})=a+bd$, where $a=-4.03 \pm 0.22$, $b=4.93  \pm 0.57 $,  is obtained.
\end{abstract}



\maketitle

\section{Introduction}
The model of random sequential adsorption (RSA) is very popular in studies of the colloidal deposition of particles (proteins, nanoparticles, viruses, biological cells etc.) on different types of substrate~\cite{Evans1993RMP}. The particles are randomly deposited on the substrate with the process being fully irreversible, without subsequent detachment or diffusion. The classical variant of this model assumes an absence of overlap of the new incoming particle with the previously deposited ones. With a large enough concentration of the deposited objects, they can form a spanning path between the opposite sides of the substrate and this concentration corresponds to the percolation threshold~\cite{Stauffer}. Finally, the jamming limit  will be reached beyond which no more objects can be adsorbed.

Very often in the simulations, the substrate is considered as a discrete space, e.g. a square lattice. Great effort has been devoted to studies regarding percolation and jamming for the RSA deposition of elongated particles, e.g., sticks~\cite{Becklehimer1992}, line segments~\cite{Leroyer1994PRB}, rigid rods, needles~\cite{Vandewalle2000epjb} or  linear $k$-mers (particles occupying $k$ adjacent sites)~\cite{Lebovka2011PRE,Tarasevich2012PRE}. For clarity,  hereinafter we shall use the term $k$-mer.
Problems with completely disordered $k$-mers~\cite{Becklehimer1992,Leroyer1994PRB,Bonnier1994,Vandewalle2000epjb,Kondrat2001PRE,Cornette2003epjb} or partially aligned $k$-mers~\cite{Longone2012PRE,Lebovka2011PRE,Tarasevich2012PRE,Romiszowski2013} have been reported over the last two decades.
Intensive studies have shown that the jamming concentration continuously decreases as the length of the $k$-mer  increases~\cite{Lebovka2011PRE}. On the other hand, the percolation threshold initially decreases and then increases with increasing value of $k$~\cite{Tarasevich2012PRE}. For a completely disordered system, a conjecture has been offered that percolation is impossible when $k$ exceeds approximately  $1.2 \times 10^4$~\cite{Tarasevich2012PRE}. Direct verification of the conjecture is very time-consuming and problematic even with a high-performance computer.

Different variants of the more general RSA models, taking account of the heterogeneity of substrates, interactions between the deposited particles and the possibility of surface diffusion have been proposed~\cite{Talbot2000,Privman2000,Senger2000,Weronski2005,Adamczyk2005}.
These models are more realistic in their description of the experimental results for colloid particle adsorption on substrates characterized by a wide spectrum of binding energies.

Very often, the real surfaces are chemically heterogeneous and contain defects~\cite{Adamson1997}, moreover, the substrates may be prepatterned~\cite{Cadilhe2007JPhysCM}. The structure of the elongated particles, e.g., carbon nanotubes, adsorbed on the substrate may also be highly heterogeneous, e.g. due to their chemical functionalization~\cite{Wepasnick2010}.
The jamming and percolation of $k$-mers on disordered (or heterogeneous) substrates with defects, or $k$-mers with defects, have also attracted great attention~\cite{Ben-Naim1994JPhysA,Lee1996JPhysA,Budinski-Petkovic2002,Kondrat2005,Kondrat2006,Cornette2003epjb,Cornette2006PLA,Cornette2011PhysA,Budinski-Petkovic2011,Budinski-Petkovic2012,Tarasevich2015PRE}.
Two models of non-ideal lattices and objects were analyzed~\cite{Tarasevich2015PRE}. In the first model, it is assumed that the initial square lattice is non-ideal and that some fraction of the sites, $d$, is occupied by point defects (impurities). The lattice sides occupied by these point defects are forbidden for deposition of the objects. In the second model, it was assumed that the square lattice is perfect and some fraction of the sites in the $k$-mers, $d$, consists of defects, i.e., is non-conducting. For both models, above some critical concentrations of defects, $d_m$, the percolation is blocked even at the jamming concentration of $k$-mers. The estimations predicted the absence of percolation even for non-defective systems when $k\gtrapprox 6 \times 10^3$~\cite{Tarasevich2015PRE}.
Integration of the above models into one generalized model in which the defects are presented both on the lattice and inside the $k$-mers looks very attractive and promising, because in the real world, both the deposited objects and the substrates are,  as a rule, non-ideal. We shall denote this model as the LK$_d$  model.

Another generalized RSA model of cooperative sequential adsorption is that  in which the adsorption probabilities are dependent on the local environment~\cite{Evans1993RMP} is of special interest. This model takes account of the presence of very strong near-neighbor (NN) lateral repulsive interactions. In the simplest case, the constraint assumes that all NN locations are empty.  For the deposition of monomers on a square lattice with complete NN exclusion, jamming is observed at a coverage $p_j=0.3641$. However, for this problem, percolation never actually occurs due to jamming. In the general case, a  restricted number of lateral contacts with previously deposited particles is allowed. In this model with partial NN exclusion (the C$_d$ model), the fraction of forbidden NN contacts, $d$, may be identified with the fraction of defects that influence the process of deposition. Note that the defect-free variant of this model ($d=0$) corresponds to the classical RSA model.

The goal of the present research is to investigate the jamming and percolation for the LK$_d$  and C$_d$ models of the deposition of linear $k$-mers on a square lattice. In the LK$_d$  model, the simultaneous presence of defects both on the substrate and in the deposited objects is assumed. For this model, the defects on the substrate influence the deposition process and the defects on the $k$-mers affect the connectivity of the system. In the C$_d$ model,  restricted number of lateral contacts with previously deposited $k$-mers is allowed  for each new $k$-mer. For this model, the fraction of forbidden NN contacts (or concentration of defects $d$) influences the deposition process.

The rest of the paper is constructed as follows. In Section~\ref{sec:methods} we describe the technical details of our simulations. Section~\ref{sec:results} presents our main findings. In Section~\ref{sec:conclusion}, we summarize the results and conclude the paper.

\section{Details of simulation\label{sec:methods}}
The percolation and jamming behavior of elongated objects on a substrate was investigated using computer simulation.
In both the LK$_d$ and C$_d$ models, we considered a discrete two-dimensional substrate (square lattice $L \times L$ sides) with periodic (toroidal) boundary conditions. The deposited objects were linear $k$-mers (particles occupying $k$ adjacent sites). Isotropic deposition was simulated, i.e., the $k$-mers being deposited in two allowed perpendicular directions (vertical or horizontal) with equal probabilities.
To extrapolate the results of the simulation to the thermodynamic limit ($L \to \infty$),  we performed a scaling analysis~\cite{Stauffer}.

\subsection{LK$_d$  model}\label{subsec:RSAdetails }
\begin{figure*}
  \centering
\includegraphics[keepaspectratio=true,clip=on,width=0.9\textwidth]{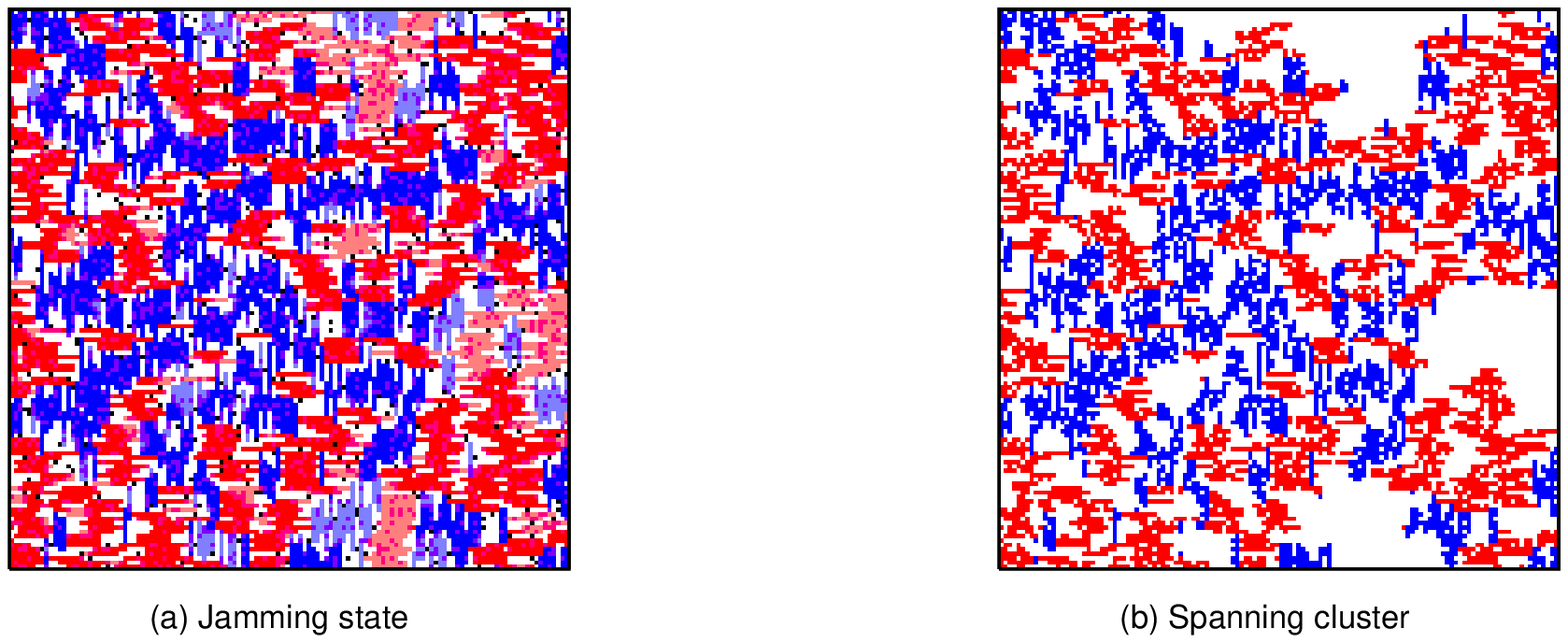}
  \caption{Example of a jamming state of isotropically deposited $k$-mers with defects on a square lattice with defects. The lattice size is $1024 \times 1024$, $k = 8$, a fragment of the lattices with $128 \times 128$ sites is shown.
  The concentration of defects on the lattice is 0.02. The concentration of defects on the $k$-mers is 0.2.
  Online: Horizontal $k$-mers are shown in red, vertical $k$-mers are shown in blue, $k$-mers belonging to the percolation cluster are shown in the same colors but with different brightness, empty sites are shown in white, defects on the lattice are shown in black, defects on the $k$-mers are shown in the tint of the color of the $k$-mer. Print: gray-scale.
  \label{fig:lattice}}
\end{figure*}

In the LK$_d$ model, the presence of defects with a concentration of $d_k$ on the deposited $k$-mers was assumed. The length of the $k$-mers, $k$, varied from 2 to 64. The lattice was also imperfect, i.e. some fraction of the lattice sites, $d_l$ was occupied by defects (impurities) (Figure~\ref{fig:lattice}). The defects hinder the adsorption of the elongated objects.

 For each given concentration of the defects on the lattice, $d_l$,
 \begin{itemize}
   \item we filled the lattice with randomly distributed point defects,
   \item then, we deposited $k$-mers up to the jamming state, using RSA,
   \item subsequently, we randomly placed defects onto the $k$-mers; the defect concentration, $d_k$, being defined as the fraction of insulating sites on the $k$-mers, i.e. the number of such defects is $d_k p L^2$,
   \item and finally, we checked whether percolation occurred.
 \end{itemize}
We filled the lattice at least 1000 times and found the probability, $R(d_k)$, that percolation would occur at given concentrations of defects on the $k$-mers.

In contrast to~\cite{Newman2001PRE}, we chose to treat spiral clusters as wrapping (percolating). We checked the percolation in two perpendicular direction and used two criteria: there is percolation in both directions (criterion AND), or there is percolation at least along one direction (OR). The abscissa of the inflection point of the curve was treated as the estimation of the critical concentration of defects for the given lattice size. We utilized a scaling relation to obtain the critical concentration at the thermodynamic limit $$
d_k(L)- d_k(\infty))\sim L^{-1/\nu_c},
$$
where $\nu_c = 4/3$ is the critical exponent~\cite{Stauffer}.
 For $k \leq 32$, we used the lattice sizes $L=100k, 200k,$ and $400k$. For $k = 64$, we used only two lattice sizes $L=100k$ and $200k$ but the scaling analyzes were performed using the two criteria (AND, OR).

\subsection{C$_d$ model}\label{subsec:CSAdetails}
In the C$_d$ model,  deposition occurs for the case when $z\leq (1 - d) z_m$  where $z_m=2(k+1)$ is the maximum numbers of contacts of the $k$-mer, and $d$ is the fraction of forbidden NN contacts. We filled up the lattice to a given concentration of $k$-mers, $p$, 1000 times and found the probability, $R(p)$, that of percolation occurring. The abscissa of the inflection point of the curve was treated as the estimation of the percolation threshold for the given lattice size. The filling fraction of the lattice by $k$-mers is defined as $p= Nk/L^2$, where $N$ is the number of $k$-mers. The value $p$ changes within $[0;p_j]$, where $p_j$ is the jamming concentration.

\begin{figure*}
  \centering
\includegraphics[keepaspectratio=true,clip=on,width=0.9\textwidth]{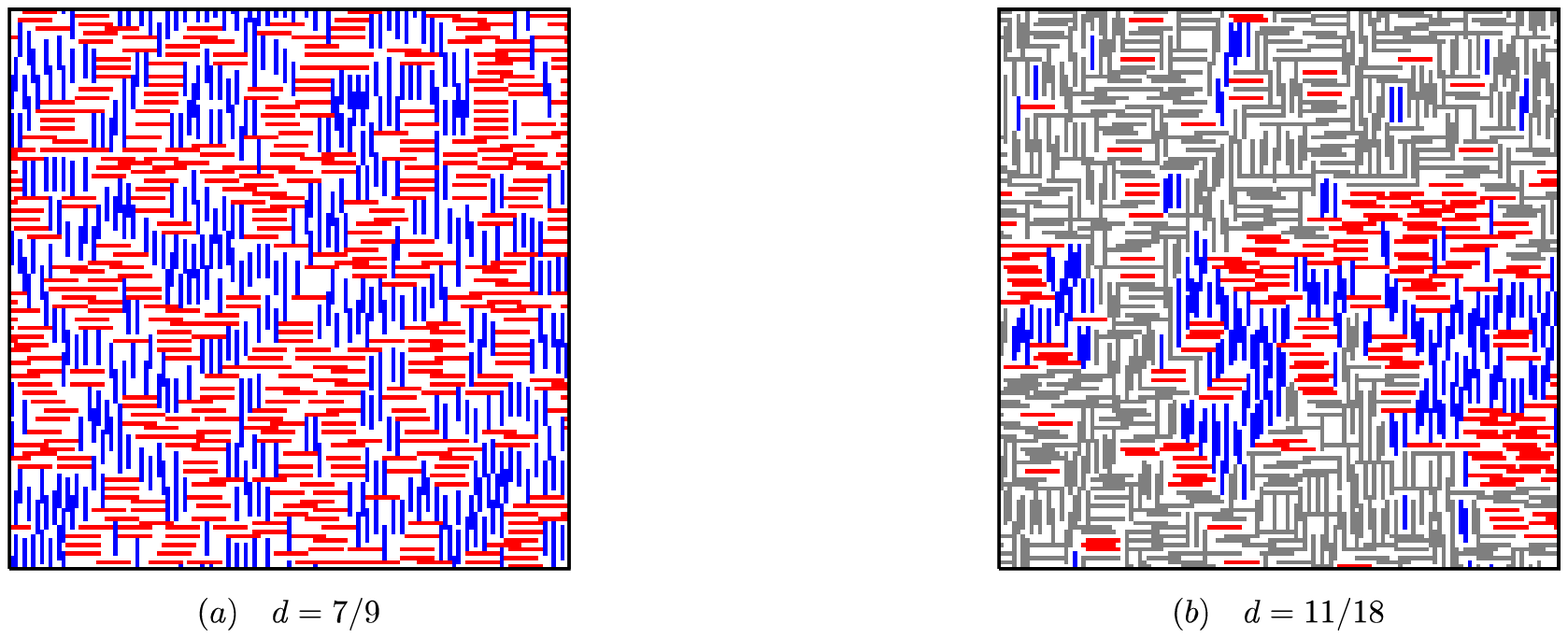}
  \caption{Jamming states for different fractions of forbidden contacts (C$_d$ model).  The lattice size is $1024 \times 1024$, $k = 8$, $z_m=18$,  fragments of the lattices with $128 \times 128$ sites are shown. Percolation is absent for large numbers of forbidden contacts, $z\geq 14$, (a) but is present at smaller numbers of forbidden contacts, $z\geq 11$, $d = 11/18$ (b). Online: Horizontal $k$-mers are shown in red, vertical $k$-mers are shown in blue, $k$-mers of the percolation cluster are shown in gray, empty sites are shown in white. Print: gray-scale.
  \label{fig:lattices}}
\end{figure*}

Figure~\ref{fig:lattices} demonstrates typical examples of the jamming states for the C$_d$ model with restricted numbers of lateral contacts for $k=8$ ($z_m=18$) and different values of $d$. Percolation is absent at a large number of forbidden contacts, $z\geq 14$ ($d=14/18$) (~\ref{fig:lattices}a) but is present for smaller number of contacts, $z\geq 11$ ($d=11/18$) (Fig.~\ref{fig:lattices}b). The decrease of lateral repulsion between the $k$-mers results in increasing  connectivity of the system and in the appearance of a percolation cluster (Fig.~\ref{fig:lattices}b).

We used several different lattice sizes ($L = 32k, 64k,128k$, and $256k$) to perform a scaling analysis and to find the jamming concentration $p_j$ and the percolation threshold $p_c$ at the thermodynamic limit ($L \to \infty$) (see, e.g.~\cite{Stauffer})
$$
p_{j,c}(L) -p_{j,c}(\infty))\sim L^{-1/\nu_{j,c}},
$$
where $\nu_{j}$ and $\nu_{c}$ are the universal critical exponents for the jamming and percolation, respectively. In two dimensions, $\nu_j = 1$~\cite{Evans1993RMP} and $\nu_c = 4/3$~\cite{Stauffer}.

\begin{figure}
  \centering
  \includegraphics[width=0.9\linewidth]{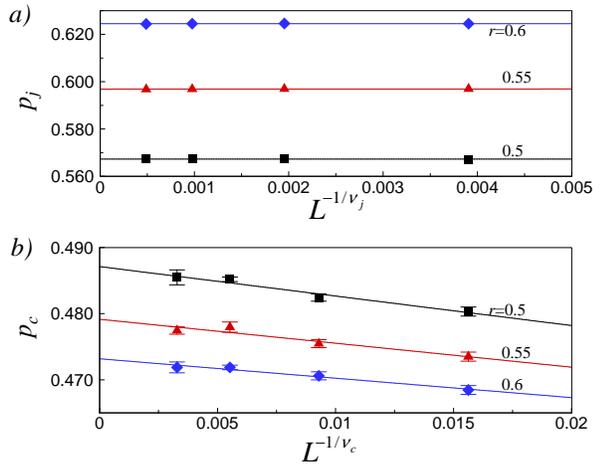}\\
  \caption{Examples of scaling dependencies for the jamming $p_j$ (a) and percolation $p_c$ (b) concentrations at different values of $r=0.5(1), 0.55(2)$, and $0.6(3)$. Here, $\nu_j = 1$ and $\nu_c = 4/3$. $k=8$. C$_d$ model.
The statistical error is smaller than the marker size when not explicitly shown.\label{fig:pjpc}}
\end{figure}

The examples of scaling are shown in Fig.~\ref{fig:pjpc}a (jamming concentration, $p_j$) and Fig.~\ref{fig:pjpc}b (percolation concentration, $p_c$).

\section{Results and Discussion\label{sec:results}}

\subsection{LK$_d$  model}\label{subsec:KdLd}
An example of a jammed state of $k$-mers with defects on a square lattice with defects is shown in Fig.~\ref{fig:lattice}a. Even at jamming filling, the percolation cluster looks rather sparse and lacy (Fig.~\ref{fig:lattice}b).

\begin{figure}
  \centering
    \includegraphics[width=\linewidth]{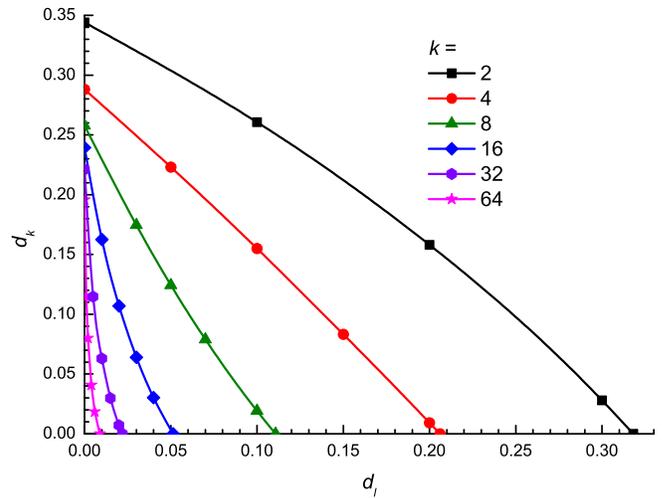}
  \caption{Phase diagram presented as the concentration of the defects on the lattice, $d_l$, versus the concentration of the defects on the $k$-mers, $d_k$. The areas below the curves correspond to the percolation states.\label{fig:phases}}
\end{figure}

\begin{figure}
  \centering
     \includegraphics[width=\linewidth]{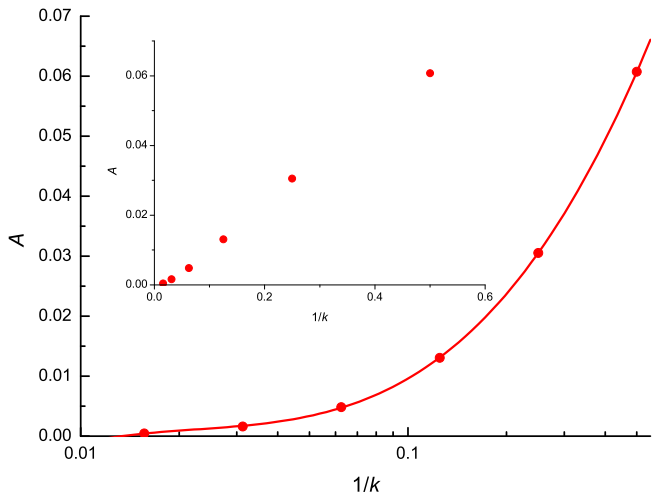}
  \caption{Area of the percolation state on the phase diagram as a function of the inverse of the length of the $k$-mers. The circles correspond to the simulation data, the curve corresponds to a fitting~\eqref{eq:fitting}\label{fig:area}}
\end{figure}

The results of the simulations are presented in Fig.~\ref{fig:phases} as phase diagrams on the plane $(d_l,d_k)$. For any given $k$, percolation can occur below the  $d_l(d_k)$ curve.  Even a small concentration of defects inhibits the deposition of long objects. The curves separating percolating and non-percolating states are convex for short objects ($k < 4$) and concave for the larger objects ($k > 4$). As the value of $k$ increases, the area of the percolating region on the phase plane $(d_l, d_k)$ decreases faster than the product of $0.5d_l d_k$  (Fig.~\ref{fig:area}). We can expect that in the case of long objects, the area will be extremely narrow and located along the axes. The result suggests that for large objects, even a very small concentration of defects can inhibit the percolation when both kinds of defects are present.

The area of the percolating state $A$ in the phase plane (Figure~\ref{fig:area}) may be well fitted by the function
\begin{equation}\label{eq:fitting}
A = b_0 + b_1 k^{-1} + b_2 k^{-2} + b_3 k^{-3},
\end{equation}
where
$b_0 = 0.1073 \pm 0.0005$,
$b_1 = 0.186 \pm 0.002$,
$b_2 = 0.111 \pm 0.002$,
$b_3 = 0.0224 \pm 0.0006$, and $R^2 = 0.9999$.
The limit of $A=0$ corresponds to the critical length of $k_{max}=74$ for the absence of percolation. This small value is unrealistic and is a result of an extrapolation based on data for rather short objects. Such extrapolation cannot be treated as reliable.

The LK$_d$  model looks very promising to describe the real systems, nevertheless, it cannot give a sufficient estimation of the critical length of the $k$-mers at which percolation is impossible.

\subsection{C$_d$ model}\label{subsec:CSA}
\begin{figure}
  \centering
  \includegraphics[width=0.9\linewidth]{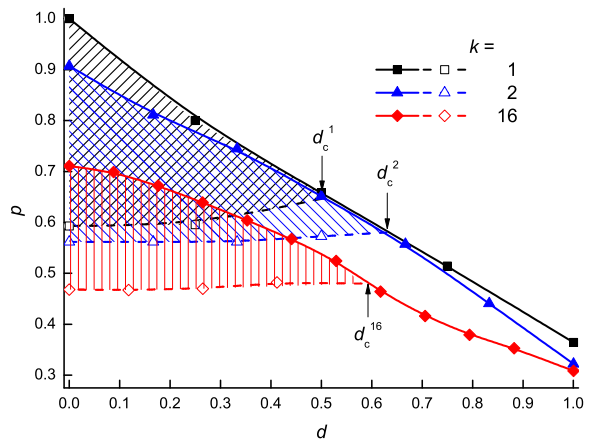}
  \caption{Examples of the jamming $p_j$ (closed symbols) and percolation $p_c$ (open symbols) concentrations versus the fractions of forbidden contacts, $d$, for $k=2$ and $k= 16$. Percolation was observed above a critical value of $d$, i.e., at $d\leq d_c$. The statistical error is smaller than the marker size.
\label{fig:pj_vs_x}}
\end{figure}

 Figure~\ref{fig:pj_vs_x} presents the jamming concentration $p_j$ versus the fraction of forbidden NN contacts, $d$, for different values of $k$-mers. The jamming concentration of $k$-mers decreases when the value of $d$ increases. It is interesting that for $k \geq2$ and $d =1$ (i.e. situation where NN is  completely excluded) the value of $p_j$ is practically independent of $k$. At larger values of $d$ the jamming concentration $p_j$ increases with~$k$.

Figure~\ref{fig:pj_vs_x} presents examples of the jamming $p_j$ (closed symbols) and percolation $p_c$ (open symbols) concentrations versus the
fractions of forbidden contacts, $d$, for $k=2$ and $k=16$. Percolation was observed above a critical value of $d$, i.e.,
at $d \leq d_c$. Increase of $d$ up to $d_c$ resulted in an increase of the percolation concentration $p_c$.
Percolation was confined between the lines $p_j(r)$ and $p_c(r)$ while at concentrations above $p_j$ percolation was suppressed by jamming.

\begin{figure}
  \centering
  \includegraphics[width=0.9\linewidth]{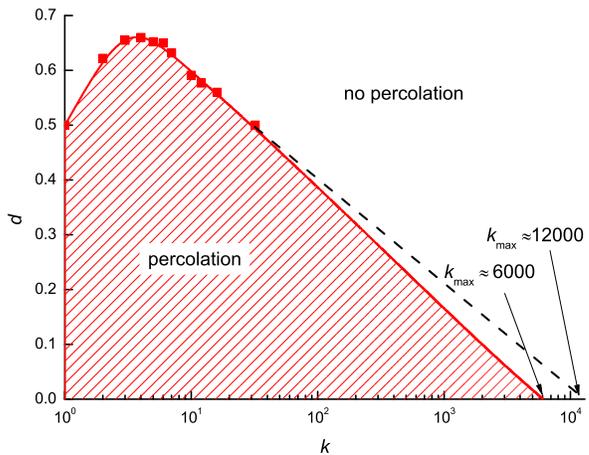}\\
  \caption{Percolation phase diagram presented as the fractions of forbidden contacts $d$ versus the length of the $k$-mers. The filled squares correspond to the critical values of $d_c$ below which percolation is present. The dashed lines correspond to the logarithmic extrapolation of $d_c(k)$ dependence to the previous estimations $k_{max}\approx 12000$ \cite{Tarasevich2012PRE}, $k_{max}\gtrapprox 6000$~\cite{Tarasevich2015PRE}. The statistical error is smaller than the marker size.  \label{fig:phase1}}
\end{figure}

Figure~\ref{fig:phase1} shows the percolation phase diagram presented as the fractions of forbidden contacts, $d$, versus the length of the $k$-mers. The critical value of $d_c$ goes through a maximum at $k\approx 4$  as the value of $k$ increases.
The asymptotic behavior $d_c$ when $k\rightarrow\infty$ may be fitted by a logarithmic function
$d_c\propto\log k$ to the limiting length $k_{max}\approx 12000$ obtained by extrapolation of the ratio $p_c/p_j$
\cite{Tarasevich2012PRE} and $k_{max}\gtrapprox 6000$ obtained for models which consider the defects only on a lattice and only in the $k$-mers~\cite{Tarasevich2015PRE}.
Thus the data presented in the present work confirms the previous estimations.
Note that the loss of percolation for very long $k$-mers is closely analogous to similar behavior observed for $k \times k$ squares \cite{Nakamura1987PRA}.

\begin{figure*}
  \centering
  \includegraphics[width=0.9\linewidth]{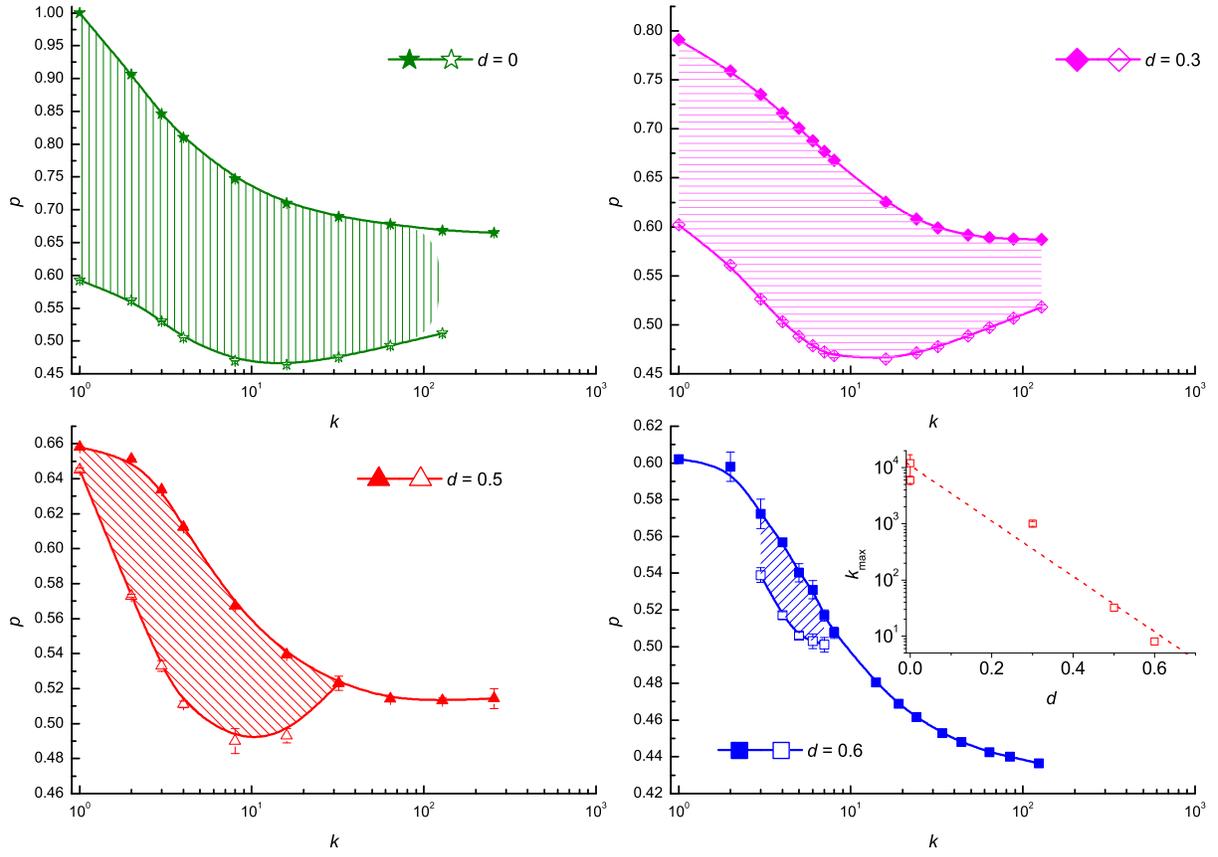}\\
  \caption{Jamming $p_j$ (filled symbols) and percolation $p_c$ (open symbols)  concentrations versus the length of $k$-mer evaluated for the fixed fractions of forbidden contacts $d=0, 0.3, 0.5$, and $0.6$. The hatched  areas between the curves $p_j(k)$ and  $p_c(k)$ correspond to the percolation regions.\label{fig:phase2}}
\end{figure*}

Figure~\ref{fig:phase2} shows examples of percolation diagrams presented as filling fraction $p$ versus the length of $k$-mer.
For a fixed fraction of forbidden contacts, $d$, the percolation region is limited by the hatched area between the curves $p_j(k)$ and  $p_c(k)$.
The jamming concentration $p_j$ continuously decreases with increasing values of $k$ whereas the percolation concentration $p_c$ goes through  a minimum. The observed behavior is quite similar to that observed for $d = 0$ \cite{Tarasevich2012PRE,Tarasevich2015PRE}. For the analyzed problem of cooperative sequential adsorption with a restricted number of lateral near-neighbor contacts,  percolation was observed in the interval $k_{min}\leq k\leq k_{max}$. The estimation gives $k_{min}=3$, $k_{max}=8$ at $d = 0.6$, and $k_{min}=1$, $k_{max}=32$ at $d = 0.5$. The inset to Fig.~\ref{fig:phase2} shows  the dependence of $k_{max}$ versus $d$. At large values of $k_{max}$ ($k_{max}\geq  10$), this dependence may be well approximated by the following logarithmic function
$$
\log(k_{max})=a+bd,
$$
where $a=4.04  \pm 0.22$, $b=-4.93  \pm 0.57 $  and the coefficient of determination is $R^2=0.9616$.

\section{Conclusion\label{sec:conclusion}}
The jamming and percolation of linear $k$-mers on a square lattice were investigated for two models. These models attempt to mimic the
processes of deposition of elongated objects on real substrates.
The first model KL$_d$ deals with the random sequential adsorption of extended inhomogeneous objects onto a substrate with preliminary deposited impurities. These impurities inhibit the deposition of the objects. The inhomogeneity of the objects means that some sites are treated as insulating. Even for not very long objects, the presence of defects of two different kinds (defects on the substrate and defects on the objects) at low concentrations prevents the percolation. The simulation suggests that for sufficiently large linear $k$-mers percolation cannot occur even on an ideal lattice.

The second model C$_d$ is based on the cooperative sequential adsorption model. This model takes into consideration interactions between the adsorbed objects.  For this model the $p_j$ and $p_c$ dependencies are controlled by the fractions of forbidden contacts $d$. The previous works investigating the case of $d=0$ (i.e., when all contacts are allowed) conjectured that the percolation of $k$-mers is impossible if their length exceeds some critical value, $k_{max}$~\cite{Kondrat2001PRE,Tarasevich2012PRE,Tarasevich2015PRE}. However, the estimations of the value of $k_{max}$ used extrapolation to very long
$k$-mers ($k \gtrsim 10^3$), as the direct simulation for such large objects was a very time-consuming task and was not very precise.
The present work for the more general C$_d$ model at $d\geq 0$ estimates $k_{max}$ more precisely for different values of $d$.
For the case of $d=0$ the data from the present work confirms the previous estimations of $k_{max}\approx 12000$ \cite{Tarasevich2012PRE} and $k_{max}\gtrapprox 6000$ \cite{Tarasevich2015PRE}. The value of $k_{max}$ decreases with increasing values of $d$.
The logarithmic dependence of the type $\log(k_{max})=a+bd$ ($k_{max} \geq 0$), where $a=4.03 \pm 0.96$, $b=-4.93  \pm 0.57 $,  is obtained ($R^2 = 0.9616$).

\section*{Acknowledgements}
The reported research is supported by the Ministry of Education and Science of the Russian Federation, Project no.~643, Russian Foundation for Basic Research, grant no.~15-02-90402 Ukr\_a, and the National Academy of Sciences of Ukraine, Project no.~43–02–15(U).

\bibliography{percolation,RSA}
\end{document}